\documentclass{PoS}
\usepackage{graphics, slashed}
\usepackage{amssymb}
\usepackage{slashed}
\usepackage[normalem]{ulem} 
\usepackage {ulem} 
\def\vecb{\bf b}

\def\veck{\bf k}
\def\vecq{\bf q}
\def\barn{\bar n}

\def\be{\begin{equation}}
\def\ee{\end{equation}}
\def\bea{\begin{eqnarray}}
\def\eea{\end{eqnarray}}
\def\dsigdt{\frac{d \sigma_{el}}{dt}}

\def\bef{\begin{figure}}
\def\ing{\includegraphics}
\def\eef{\end{figure}}
\title{Infrared Gluon Resummation and   pp total cross-sections}

\ShortTitle{Dissecting elastic pp scattering, the optical point}

\author{\speaker{Giulia Pancheri}\thanks{ Part of this work was done while a visitor at MIT-CTP}\\
        INFN Frascati National Laboratories, Frascati, Italy\\
        E-mail: \email{giulia.pancheri@lnf.infn.it}}

\author{Daniel A. Fagundes\\
       Instituto de F\'{i}sica
Gleb Wataghin, Universidade Estadual de Campinas, UNICAMP, 13083-859,
Campinas-SP, Brazil\\
      E-mail: \email{fagundes@ifi.unicamp.br}}
\author{Agnes Grau\\
      Departamento de Fisica Teorica y del Cosmos and CAFPE,
Universidad de Granada, \\18071 Granada, Spain//
      E-mail: \email{igrau@ugr.es}}
      \author{Olga Shekovtsova,\\
      Institute of Nuclear Physics PAN
ul. Radzikowskiego 152
31-342 Krakow, Poland//
E-mail:\email{olga.shekhovtsova@ifj.edu.pl}}
      \author{Yogendra N. Srivastava\\
       Dipartimento di Fisica, University of Perugia, Perugia, Italy\\
      E-mail: \email{yogendra.srivastava@pg.infn.it}}
\abstract{
 
We address here the problem of describing both the total \& the elastic proton-proton cross-section, through        the four outstanding features  of hadron scattering: (i) the optical point; (ii) the forward peak, (iii) the dip \& (iv) the subsequent descent at larger momentum transfers. These issues are discussed through an eikonal  model for the elastic amplitude where the matter distribution in impact parameter space is given by resummed soft gluons down into the infrared (IR) region. The asymptotic growth of the total cross-section is    obtained in a mini-jet model and the taming (saturation) at high energies  is related to confinement realized here through an IR singular strong coupling constant $\alpha_s(Q^2)$. We present an ansatz that links the IR singularity of $\alpha_s(Q^2)$ to that of asymptotic freedom (AF) (at lowest order). Through this model, we illustrate the problems that arise in a {\it generic} one-channel eikonal model employed for a description of the measured differential elastic cross-section at LHC7.  }
 
\FullConference{Photon 2013,\\
		20-24 May 2013\\
		Paris, France }

\begin{document}

\section{Introduction}
The momentum transfer ($t$) dependence of the elastic differential cross-section,  recently measured by the TOTEM Collaboration
 \cite{totem},  can be dissected into  the  following elemental components:
\begin{itemize}
\item{\it the optical point}:, i.e. the value at $t=0$  which is proportional to the square of the total cross-section, 
modulo a small correction due to $\rho(s)^2$, where $\rho$
 is the ratio of the real to imaginary part of the elastic amplitude at $t=0$,
 \item{\it the Coulomb region}: which carries a $t$-dependence $\sim F_{p}^{4}(t)/|t|^{2}$, 
 through the well-known Rutherford scattering cross section, represents a key piece to determine 
 $\rho(s)$ in the interference region with hadronic amplitude at small $t$. 
\item{\it the forward peak}, i.e.  the decrease for small $t$ values  is well described by 
an exponential in $t$, i.e. a gaussian fall off in the Fourier transform space, $\vecb$-space,
\item{\it the dip-bump}, namely the occurrence of   a sharp dip    for $t_D=-t_{dip}$ with   
$t_D$  an energy dependent value, which can be seen to  satisfy   a new geometric scaling law,  
$\tau_{geo}=t_D \sqrt{\sigma
_{elastic}\sigma_{total}}=constant$ \cite{Pancheri:2014},
 \item{\it the tail}, the region after the dip-bump, where the cross-section starts decreasing 
 once more with a behaviour equally well described by an exponential in $t$ or by a power law, 
 i.e. $|t|^{-n}$ with $n\lesssim 8$ \cite{totem,DL2013}. 
\end{itemize}
{ In what follows we shall address the optical point question and the behavior around the dip, including the scaling properties,
with a mini-jet model inclusive of IR gluon resummation \cite{our99,our2005}
( labelled as BN for reasons to be explained ). We shall start with the forward peak, as a way to introduce the BN model as well. }
\section{The optical point, i.e. the total cross-section}
With the  notation:
\bea
\sigma_{total}=4 \pi \Im m \mathcal{F}(s,0) \nonumber \\
\dsigdt= \pi | i\Im m \mathcal{F}(s,t) + \Re e \mathcal{F}(s,t)|^2 \\ 
\eea
the value of the differential elastic cross-section at the optical point, i.e. at $t=0$, 
can be written as
\bea
\dsigdt|_{t=0}= \pi |\Im m \mathcal{F}(s,0)|^2 (1+\rho(s)^2)= [\frac{\sigma_{total}^2}{16 \pi}] [1+\rho(s)^2] \\
\rho(s)=\frac{\Re e \mathcal{F}(s,0)}{\Im m \mathcal{F}(s,0 } 
\eea
Thus, to reproduce the optical point value, we need a good model for the total cross-section. 
Let us begin with  a simple  single-eikonal model, with a { purely  imaginary} amplitude, i.e.,
\be
\mathcal{F}(s,t)=i\int [\frac{d^2 \vecb}{2\pi}] e^{i\vecb \cdot \vecq} [1-e^{-\barn (b,s)/2}] \label{eq:eik1}
\ee
with $-q^2=t$. We have seen \cite{our2011} that in { one channel}  eikonal models such as the above,   
the purely imaginary eikonal function corresponds to the average number $\bar {n}$
of independent collisions in impact parameter space. At high energy, such collisions are mostly 
describable through QCD parton-parton collisions. Postulating factorization between the energy 
dependence and the impact momentum variable, and an empirical separation between soft and 
hard, both  reasonable approximations at high energy, one can write 
\be
\barn=\barn_{soft}(b,s) +\barn_{hard}(b,s)\label{eq:eik2}
\ee
In our mini-jet model \cite{our99,our2005,PLB2008}, we parametrize the low energy behaviour, 
namely up to $\sqrt{s}\approx 5\ GeV$, while making the hypothesis that the rise is due to QCD processes, 
namely parton-parton scattering, resulting in so called mini-jets. In such models, one writes:
 \bea
 \barn_{soft}(b,s)=A_{soft}(b,s)\sigma_{soft}(s) \label{eq:eik3}\\
 \barn_{hard}(b,s)=A_{hard}(b,s)\sigma_{mini-jet}(s) \label{eq:eik4}
 \eea
  To determine the impact parameter dependence, we apply two different procedures to the soft and to 
  the hard term. Since our aim, at present, is to apply QCD to the rising part of the cross-section,  
  for the soft term our model uses an impact parameter distribution obtained from the convolution of the 
  electromagnetic form factors of the colliding particles. For the hard,  mini-jet,  term, we introduce the 
  Fourier transform of the Soft Gluon Resummation expression, namely
 \bea
 A_{hard}(b,s)\equiv A_{BN}(b,s)=\mathcal{N} e^{-h(b,s)}\\
 h(b,s)=\frac{8}{3 \pi^2} \int_0^{qmax}  d^2 {\veck}_t \frac{\alpha_s(k_t^2)}{k_t^2}[1-e^{-i\vecb \cdot \veck_t}] \ln(2 q_{max}/k_t)\label{eq:hdb}
 \eea 
where the subscript $BN$ refers to the fact that in this model all values of $k_t$ contribute to the integral, 
just as in the original  Bloch and Nordsieck discussion of the infrared divergence in electrodynamics \cite{BN}. 
The upper limit of the single $k_t$ integration, is called $q_{max}$  and depends on energy and the  
specific parametrization of the Parton Density Functions (PDFs)  used in the mini jet cross-sections 
\cite{our99,Corsetti}. We have commented about the function $h(b,s)$ in our publications, here we recall 
our ansatz:   when 
 $k_t<\Lambda$  we introduce  a singular   effective quark-gluon coupling, albeit the final expression is integrable, namely
 \be
 \alpha_s(k_t^2)\simeq (\frac{k_t}{\Lambda})^{-2p},  \ \ \ \  k_t^2<<\Lambda^2\label{eq:alphas}
 \ee
 with $p<1$ for the integral of Eq. ~(\ref{eq:hdb}) to exist. { In \cite{our99} we had made the phenomenological approximation}
 \bea
  \alpha_s(k_t^2)=\frac{p}
  {b_0 \ln[ 1+p(\frac{k_t^2}{\Lambda^2})^{p}] } \label{eq:alphasemp} \\
    \alpha_s(k_t^2)\to \frac{1}{b_0} (\frac{k_t}{\Lambda})^{-2p} \ \ \ \ k_t^2<<\Lambda^2 \label{eq:alphas1999paper}\\
      \alpha_s(k_t^2)\to \alpha_s^{AF}(k_t^2)=
      \frac{1}
      {b_0 
      \ln [
      \frac{k_t^2}{\Lambda^2}
      ]
      }
       \ \ \ \  k_t^2>>\Lambda^2
 \eea
with $b_0=(33-2N_f)/12 \pi$. A simpler and more interesting proposal would be to identify the singularity parameter  
$p$ with the LO coefficient of the $\beta-$ function, $b_0$,  through the following simple rewriting of the QCD 
Asymptotic Freedom  expression for $\alpha_s$, i.e.  since 
 \be
 \alpha_s^{AF}(Q^2)=\frac{1}{\ln[(\frac{Q^2}{\Lambda^2})^{b_0}]}
 \ee
  we propose to substitute the above with   an  effective  expression for the strong coupling constant  given by
 \be
 \label{BN}
 \alpha_s^{BN}(Q^2)=\frac{1}{\ln[1+(\frac{Q^2}{\Lambda^2})^{b_0}]}\label{eq:alphasBN}
 \ee
{Just as the empirically proposed expression in Eq.~(\ref{eq:alphasemp}),  
the expression in Eq.(\ref{BN}) can then be used in the entire region of interest for resummation,    
extended  safely down to $k_t=0$. The  suffix {\it BN} is used to 
indicate precisely this, namely that it can be used down into the infrared region. The advantage 
of this expression is that it does not require introduction of an extra parameter $p$:  
the behavior from $Q^2=0$ to $Q^2\to \infty$ is dictated only by  the anomalous dimension factor.}

\begin{figure}[htb]
\centering
\resizebox{0.6 \textwidth}{!}{
\includegraphics{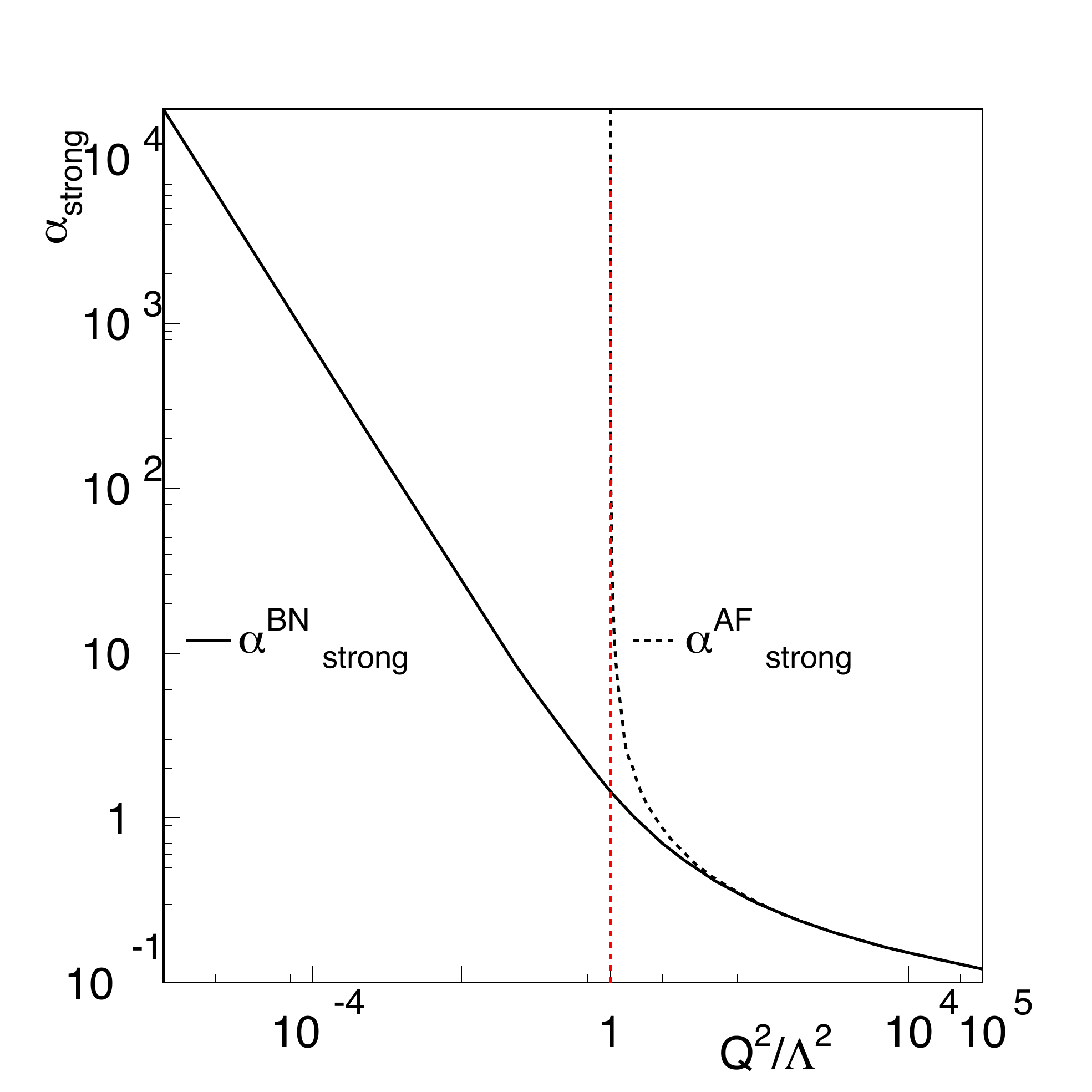}}
\caption{The asymptotic freedom expression for the strong coupling constant and our proposed expression,  
plotted as a function of $Q^2/\Lambda^2$, for the case $N_f=3$.}
\label{fig:alphasBN}
\end{figure}

The above proposal will be further elaborated in a separate publication. Here, it suffices to say that the 
anomalous dimension factor $b_0<1$, and thus $\alpha_s^{BN}$ satisfies  the condition for integrability 
of our resummation function $h(b,s)$ in Eq.~(\ref{eq:hdb}). In Fig. ~\ref{fig:alphasBN} we show 
the behaviour of the  proposed expression Eq.~(\ref{eq:alphasBN}), for the case $N_f=3$, as a function 
of  $Q^2/\Lambda^2$.

{Notice however that the numerical results we are presenting, and have presented so far, will still use the 
empirical  expression of Eq. ~(\ref{eq:alphasemp}), and that the parameter $p$  has so far been determined on purely phenomenological grounds.
 As we shall see,  values for a good representation 
of the total cross-section,  { are such that }   $p\simeq 0.6\div 0.75$ depending on the PDF's used, while   $b_0$ varies between 
0.56 and 0.77, for $N_f=6,\ 2$, respectively. The ansatz of Eq.(\ref{eq:alphasBN}) is based on the LO expression 
for $\alpha_s$, of course, and a closer correspondence needs further study.}

{The  behavior of Eq. (\ref{eq:alphas}) introduces a cut-off in $\vecb$-space \cite{ourfroissart} such that 
asymptotically the total cross-section behaves as}
 \be
 \sigma_{total}\sim [\epsilon \ln s]^{1/p}\label{eq:froissart}
 \ee
 where $\epsilon \simeq 0.3-0.4 $  corresponds to  the asymptotic behavior of the mini-jet cross-section, i.e. 
 $\sigma_{mini-jet}\sim s^\epsilon$ as $\sqrt{s}\rightarrow \infty$. The energy rise predicted by this  
 model depends on a parameter set consisting of the singularity parameter  $p$, the minimum 
 mini-jet transverse momentum $p_{tmin}$ needed to avoid the Rutherford singularity in the 
 parton-parton cross-section, and the PDFs, chosen to be at Leading Order (LO)  to avoid double counting 
 with soft resummation contributions. {  Using different PDF sets,  GRV  \cite{allGRV} and MRST \cite{MRST}, 
 we show  in Fig. ~\ref{fig:sigtot} expectations for both   total and inelastic cross-sections from \cite{our2011}.
 \begin{figure}[ht]
 \centering
\includegraphics[width=0.8\textwidth]{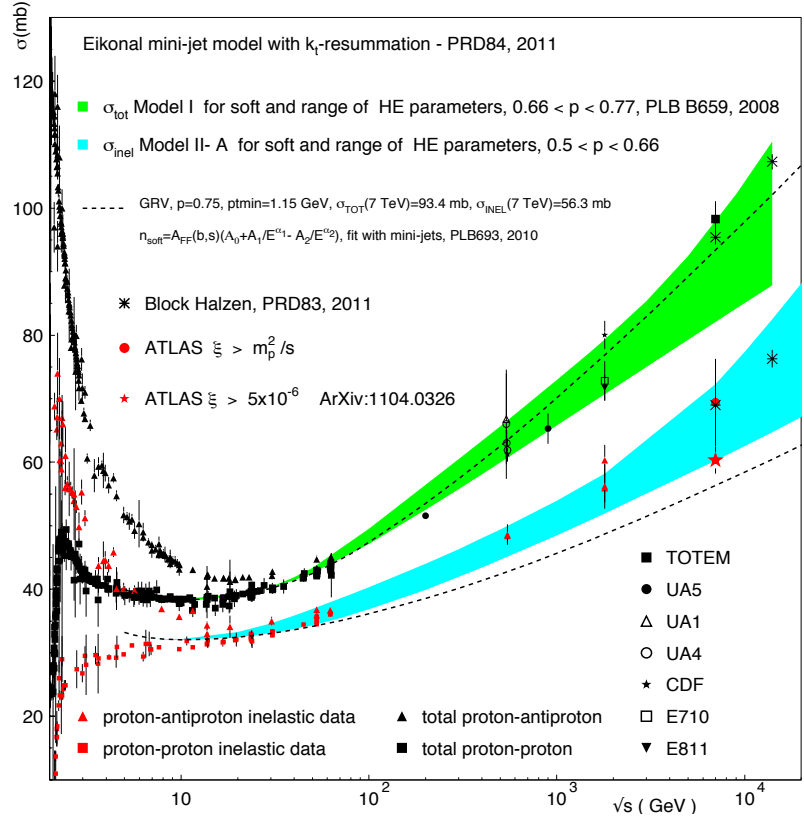}
\vspace{-3cm}
 \caption{Total and inelastic pp cross-section using the  BN model
  from \cite{our2011}.}
  \label{fig:sigtot}
 \end{figure}
Thus the  model is now well defined and can be applied to study the elastic differential cross-section. 
\section{The elastic, total and  differential, cross-section in the BN model }
\subsection{Total cross-sections, elastic and inelastic}
{ We now apply the mini-jet model with Soft Gluon Resummation (SGR) as described in the first section to study   the integrated  elastic 
and differential $pp$ cross-sections.
As pointed out in \cite{our2011}, in  
a one-channel eikonal model,  the division between elastic and inelastic contributions 
is {\it misguided }, namely   the expressions}
\bea
\sigma_{elastic}=\int d^2 {\bf b}|1-e^{-\barn(s,b)/2}|^2 \label{eq:sigel}\\
\sigma_{inel}=\int d^2 {\bf b}[1-e^{-\barn(s,b)}]\label{eq:siginel}\\
\sigma_{total}=2\int d^2 {\bf b}[1-e^{-\barn(s,b)/2}]\label{eq:sigtot}
\eea
 do not include all inelastic processes into $\sigma_{inel}$, but only  uncorrelated inelastic  ones. 
If one uses the above simple formalism, it follows  that diffractive, or correlated processes, are actually  
counted into $\sigma_{elastic}=\sigma_{total}-\sigma_{inel}$. Two or 
many-channel formalism and models for the diffractive part of the inelastic cross-section, have been   
invoked to solve this difficulty. 
We 
seek here a different 
phenomenological {\it way out}.

Such a phenomenological way out, barring the introduction of more parameters, is  to use  
different values for the singularity parameter $p$ for the elastic and the total  
cross-section. For illustration, we can use  a representative parameter set such as, for instance, 
$\{GRV,p_{tmin}=1.15\ GeV\}$. This set  allows to describe  $\sigma_{total}$ with  $p=3/4$, while  
$\sigma_{elastic}$ can be obtained 
with a different value: $p=5/6$.  {  Similarly, using for  densities the MRST72 set,
 we find that the 
 elastic cross-section data need a higher p-value than the total 
cross-section data.}
Notice that changing the PDFs and/or $p_{tmin}$, which are correlated 
parameters, will also slightly change the $p$ values, but
what matters in the discussion here is that the  $p$ value describing $\sigma_{elastic}$ is larger 
for the elastic cross-section. { As long as $\sigma_{elastic}$ is not rising at the same rate as $\sigma_{total}$, namely as long as ${\cal R}_{el}=\sigma_{el}/\sigma_{tot}\neq 1/2$, this is an obvious consequence of Eq.~(\ref{eq:froissart}), but }
if  the ansatz of Eq. ~(\ref{eq:alphasBN}) is correct, this  matches with its naive understanding: 
 in the elastic cross-section  only two flavors
  are present, and $p=b_0(N_f=2)=0.77$, 
while in the total, in principle,  all   flavors can be produced and   $p=b_0(N_f=6)=0.56$. 
 \subsection{The elastic differential cross-section}
Leaving temporarily aside the above last considerations, we  now turn to the differential elastic cross-section. 

We start with the straightforward application of Eqs.(\ref{eq:eik1}), (\ref{eq:eik2}), (\ref{eq:eik3}), (\ref{eq:eik4}) 
with a purely imaginary eikonal function, i.e. BN model
 and the  parametrization  including  soft part as in \cite{our2005}, which we define {\it a' la Regge}, i.e. 
 $p_{tmin}=1.15\ GeV,\ GRV,\ p= 3/4, \sigma_0=48\ mb$. The result is shown in the left hand plot of 
 Fig.~\ref{fig:bnelastic}  and is obviously unrealistic. 
{Apart from  the dip position, which corresponds to the first zero of the imaginary part of the amplitude 
and which occurs too early, the shape is the one of a blask disk (BD) model, with lots of dips and bumps, 
and they need to be corrected. The right hand plot shows the effect of adding a real part, following 
Martin's prescription \cite{martin-on-real}, namely}
 \bea A(s,q) = i \int bdb (1-e^{-\chi(b,s)})J_0(bq)\nonumber\\
+ \rho(s)\int bdb
(1-e^{-\chi(b,s)})[J_0(bq)-\frac{qb}{2}J_1(qb)]
\end{eqnarray}
which leads to
\begin{equation}
\frac{d\sigma}{dt}= \pi\{I_0^2+\rho^2[I_0-\frac{\sqrt{-t}}{2}I_1]^2\}  \label{eq:bnwithreal}
\end{equation}
with
\begin{eqnarray}
I_0=\int bdb (1-e^{-\chi(b,s)})J_0(qb)\\   
I_1=\int b^2 db (1-e^{-\chi(b,s)})J_1(qb) 
\end{eqnarray}

   \bef
 \ing[height=5cm,width=8cm]{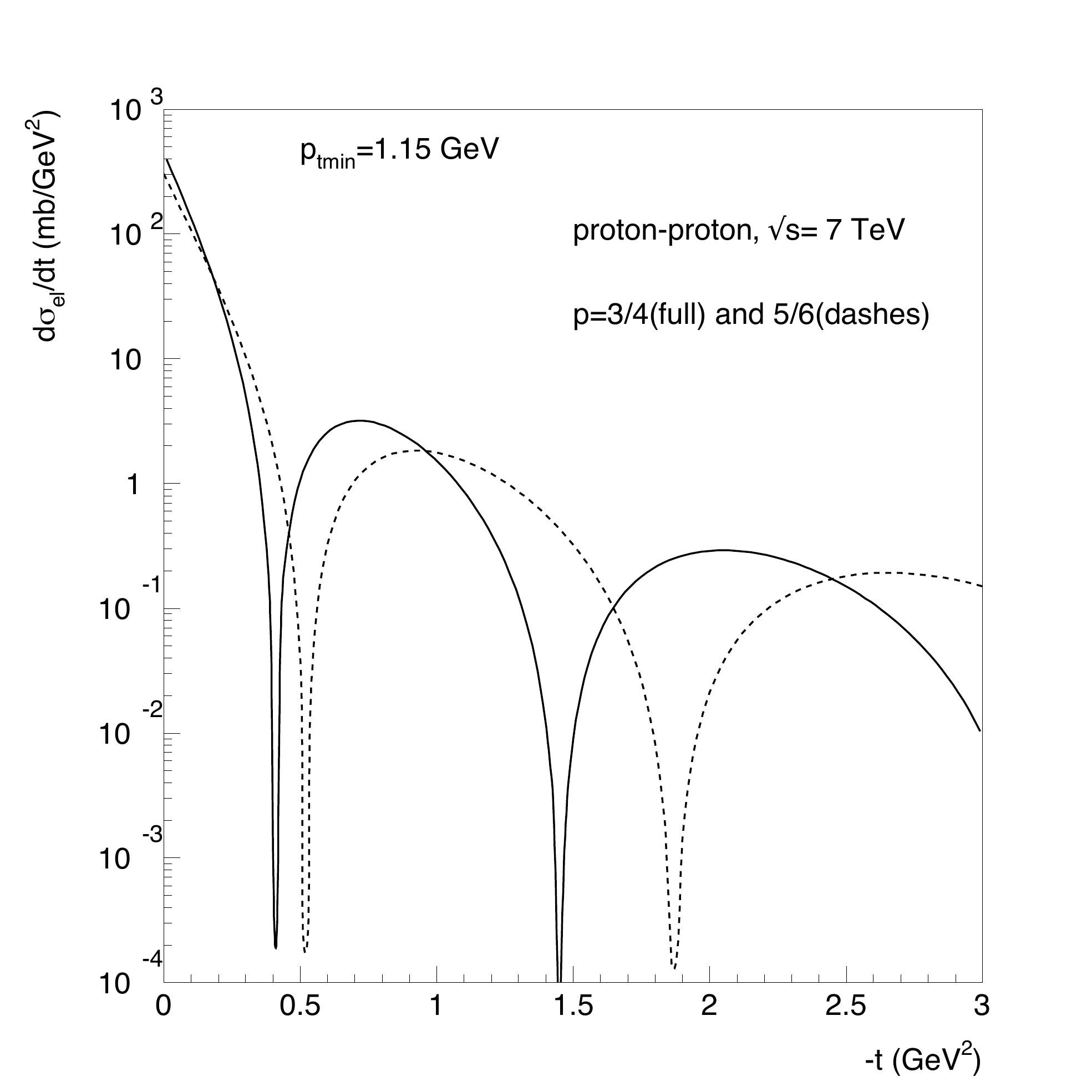}
  \ing[height=5cm,width=8cm]{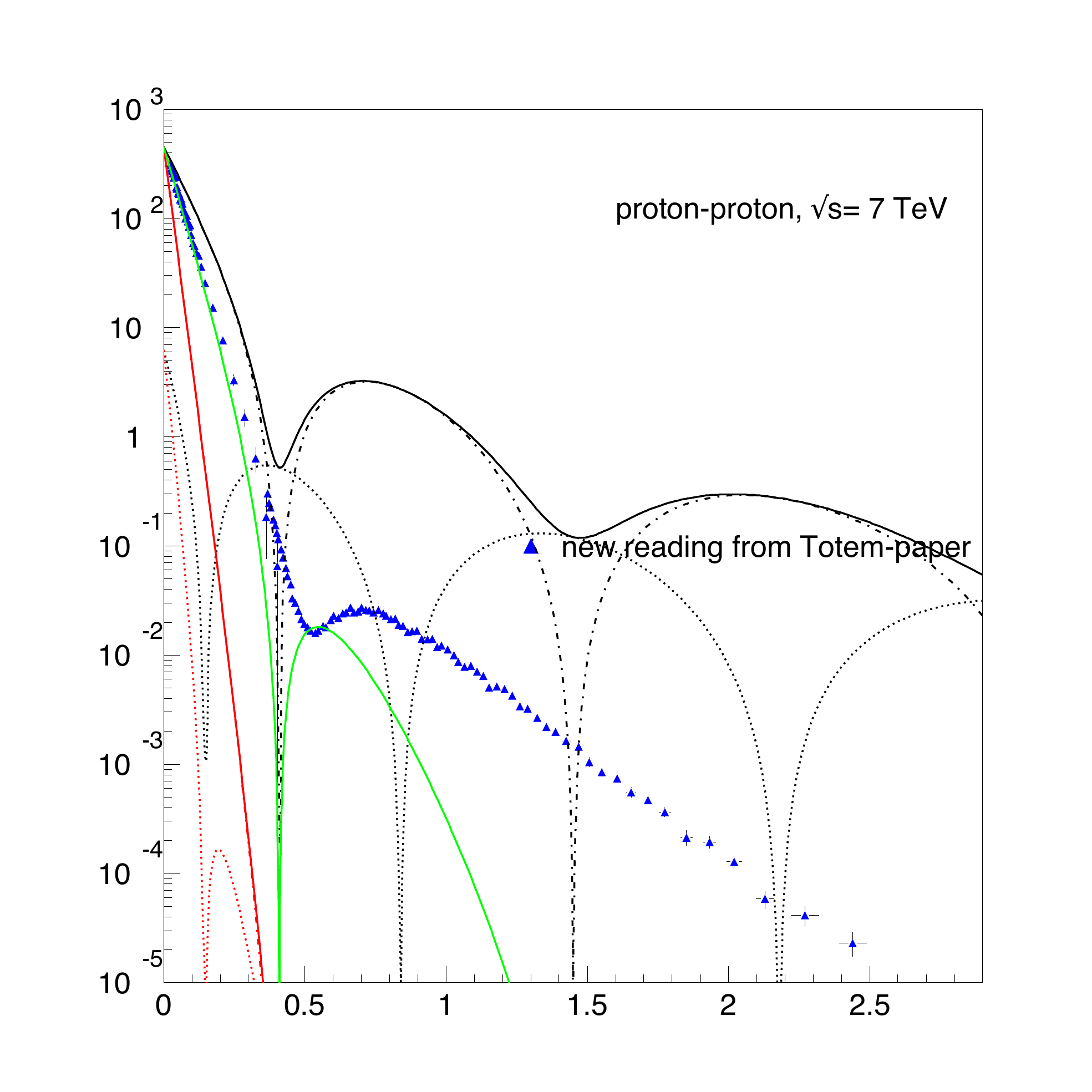}
  \caption{
 The BN model applied to the differential elastic cross-section data from TOTEM for  $\sqrt{s}=7\ TeV$. 
 At left the results of model with a purely imaginary part, at right {(full black line) with a real part (dots) added to the imaginary part (dashes) following   Martin's prescription, as indicated.   Red and green lines correspond to multiplication by a rescattering factor, as described later.}}
  \label{fig:bnelastic}
 \eef
  
Two observations are in order concerning the $t$  behavior, the first is that  at small $t$, 
the decrease is not an exponential and the cross-section  is not decreasing fast enough, 
and the second is that the dip is not reproduced at all, neither in position nor in value.  
{ As mentioned,} at very high energy,  the shape of the curve reproduces  a black disk model  as one can 
see from  the behavior of the amplitude in $b$-space, Fig. ~\ref{fig:bnsigint}.
\bef
\centering
\includegraphics[width=8cm,height=8cm]{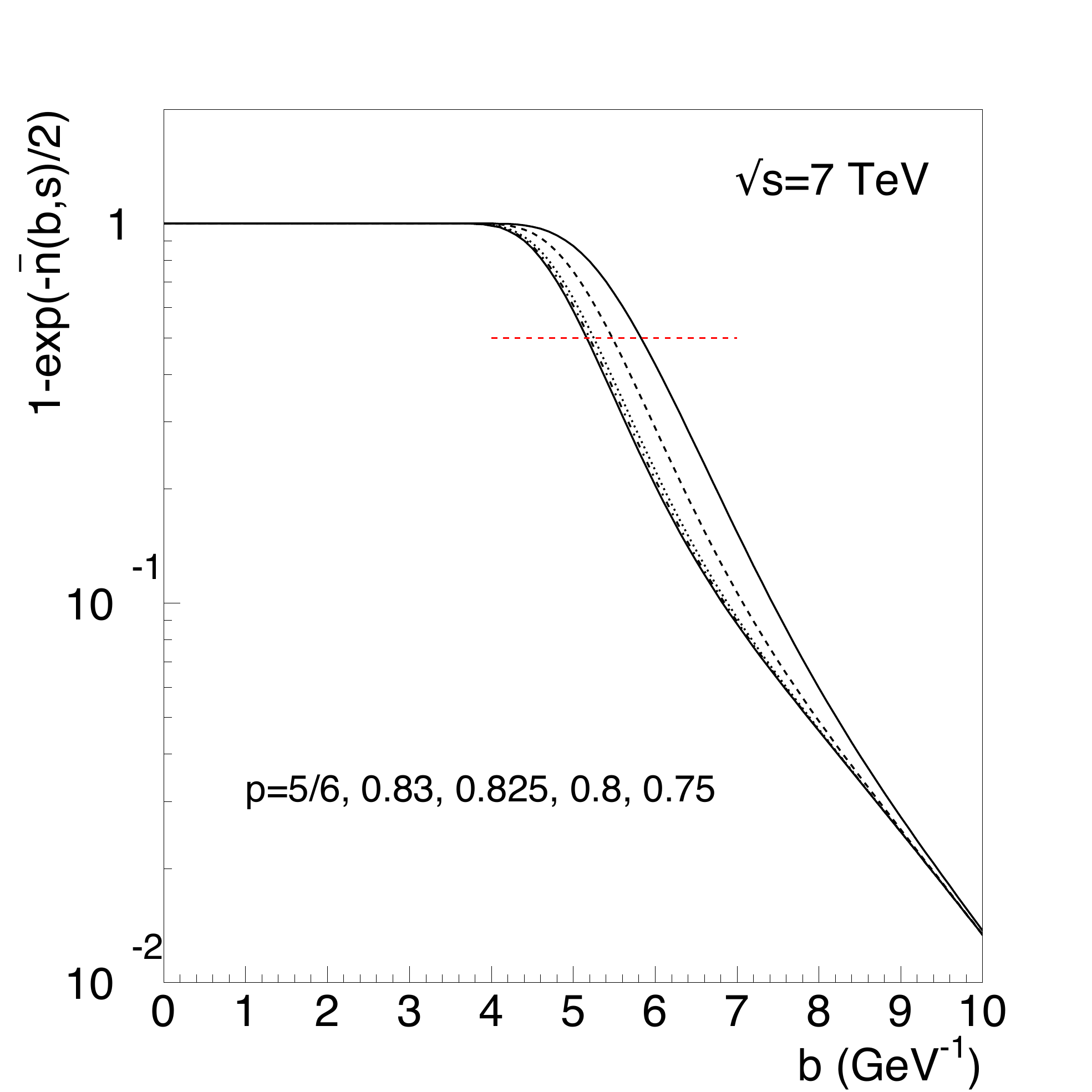}
\caption{The BN amplitude in $b$-space, for different values of the singularity parameter $p$. 
The dashed red line indicates the values when $1-exp[-\barn/2]=1/2$. Higher $p$ values 
correspond to innermost curves.}
\label{fig:bnsigint}
\eef
for a typical  parametrization with GRV densities,  $p_{tmin}=1.15\ GeV$ { and different $p$ values}
 at $\sqrt{s}=7\ TeV$. 
The red dashed line indicates the $b$-value at which { the amplitude in $b$-space is such that } ${\mathcal A}(b_{max},s)=1/2$. Then, 
one can approximate   the amplitude
  with a black disk of radius $b_{max}$, so that  
\be
A^{BD}_{el}(s,t) \approx b_{max}^{2} \left[\frac{J_{1}( b_{max}\sqrt{-t})}{ b_{max}\sqrt{-t}} \right] 
\ee 
and  the dip occurs at the first zero of the $J_1$,  located at $b_{max}\sqrt{-t_{dip}} \simeq 3.83$, leading to  $-t_{dip}\simeq14.7 /b_{max}^2$.  
One sees that $b_{max}$ at a given energy depends on the value of the singularity parameter 
$p$: larger $p$ means more suppression, and smaller $b_{max}$.  
Thus we make 
a general point of physical relevance { which follows directly from our ansatz of Eq.~(\ref{eq:alphas}):}
${\rm both}\ b_{max}\uparrow$ {\rm and}\ $\sigma_{tot}\uparrow$ as $p\downarrow$. 

The BN model corresponds to initial state emission of soft gluons from intermediate parton-parton scattering. 
A final rescattering has to be expected for  the elastic cross-section at $t\ne 0$. { In a model such as the one 
we are discussing, the only scale present is $b_{max}$ and   to reproduce the forward peak, we apply 
to the amplitude  a rescattering factor given by 
\be
{\cal S}(s,t)=e^{t b^2_{max}(s,t)/4 }\ \ \ \ \ \ \ {\cal S}(s,0)=1 \label{eq:slope}
\ee
which corresponds to
 $b_{max}(s,t)$ being  the  value beyond which the amplitude in b-space falls below $1/2$.  
 {It  is quite possible that $b_{max}$ depends on the momentum transfer $q^2$, with $b_{max}(s,t_1)>b_{max}(s,t_2)$ 
 for $-t_1<-t_2$, but in the following we shall just consider it as a constant in $t$, { namely we shall put $b_{max}\equiv b_{max}(s,0)$}.
The numerical value of $b_{max}$ is not well defined, since we could have chosen it to be the   
value where ${\mathcal A}(s,b)= [1-1/e]$ as, for instance in \cite{Block} 
and also because  the BD approximation is not exact yet, at LHC7. }Thus, the factor to be applied is here 
determined by trial and error.  Multiplying the cross-section by the factor ${\cal S}(s,t)$ with 
$4.8<b_{max}< 6\ GeV^{-1}$ gives the results shown in Fig. ~\ref{figs:bn-bmax}, obtained with mini-jets calculated with  both   MRST72 and GRV PDFs.
 \bef
 \ing[height=8cm,width=8cm]{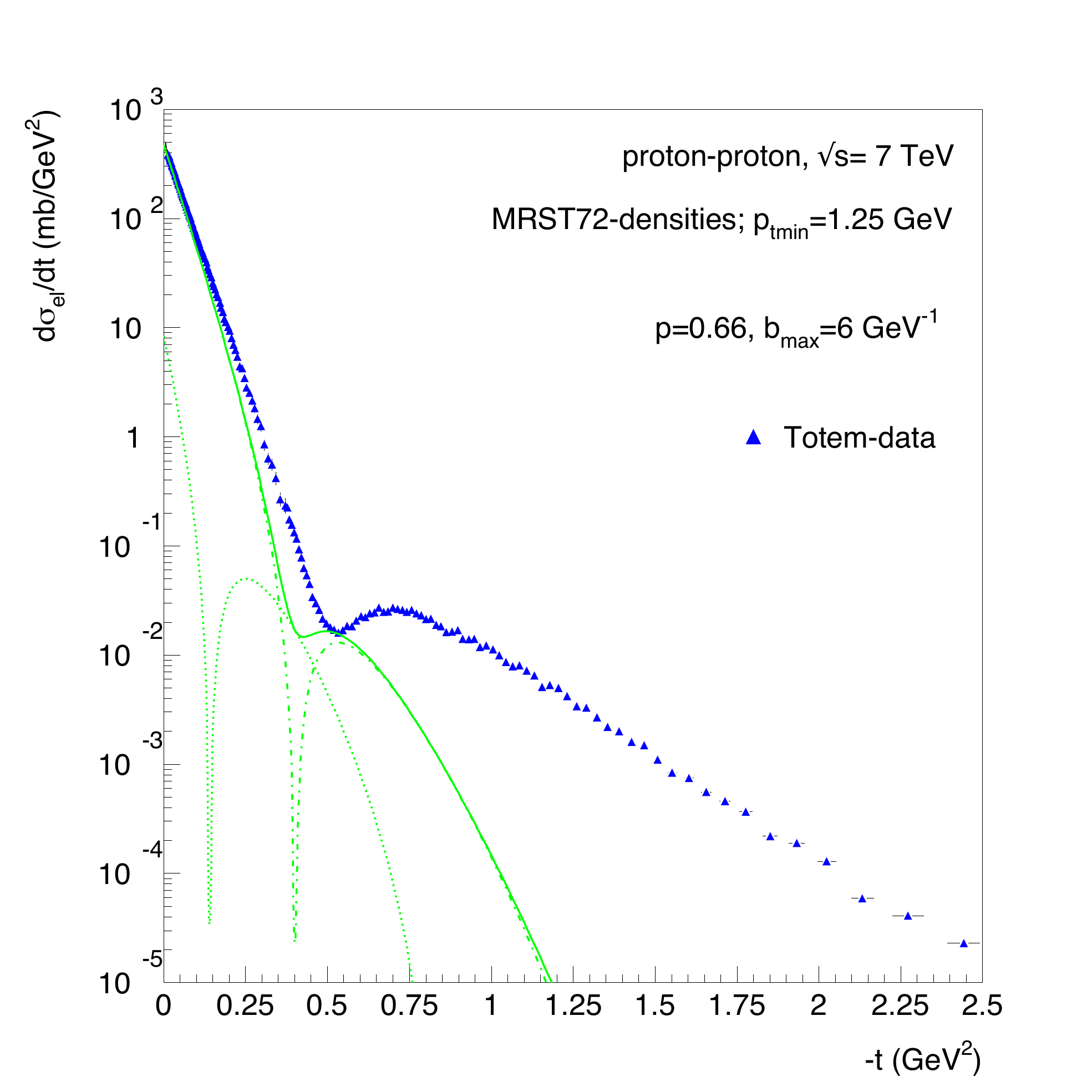}
\ing[height=8cm,width=8cm]{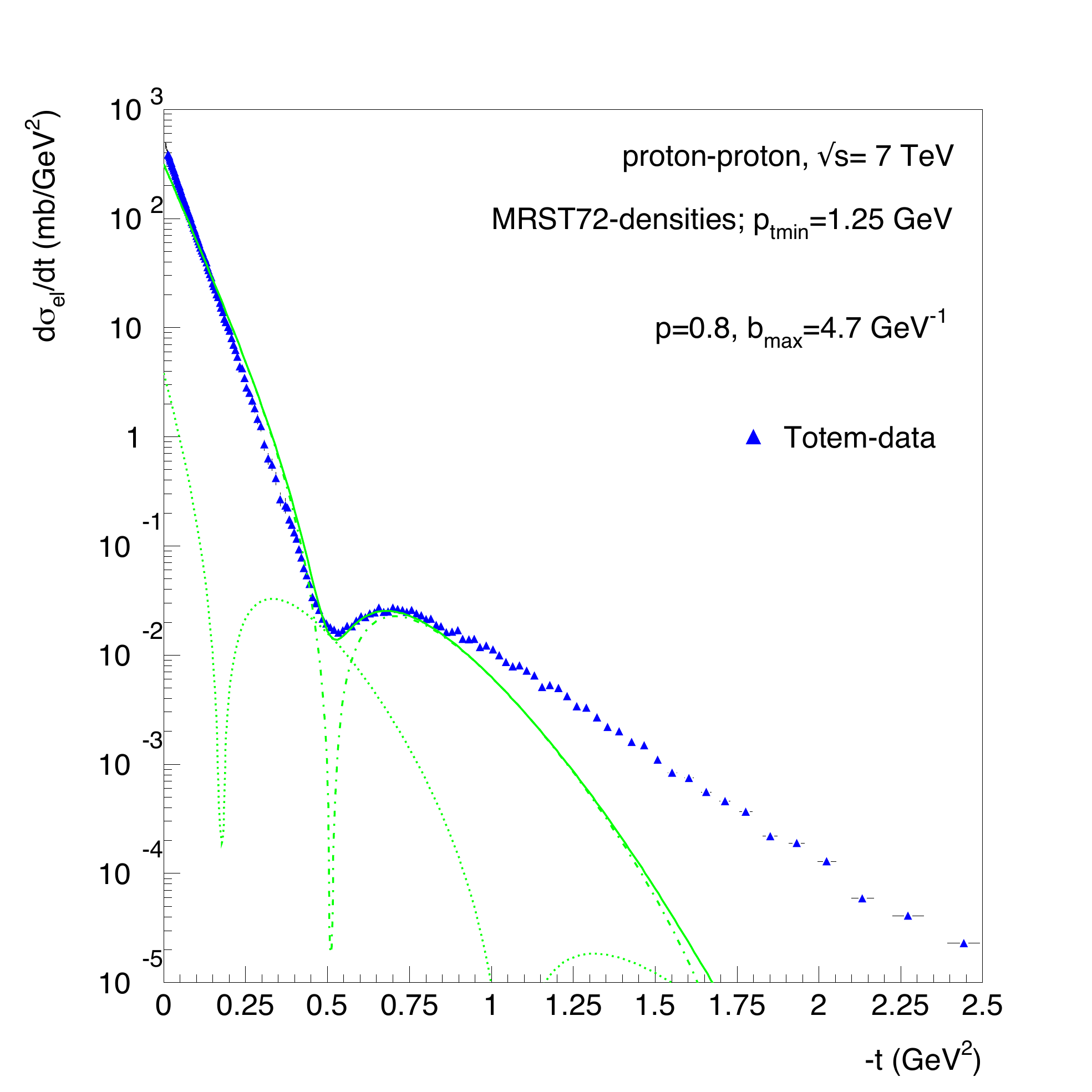}
\ing[height=8cm,width=8cm]{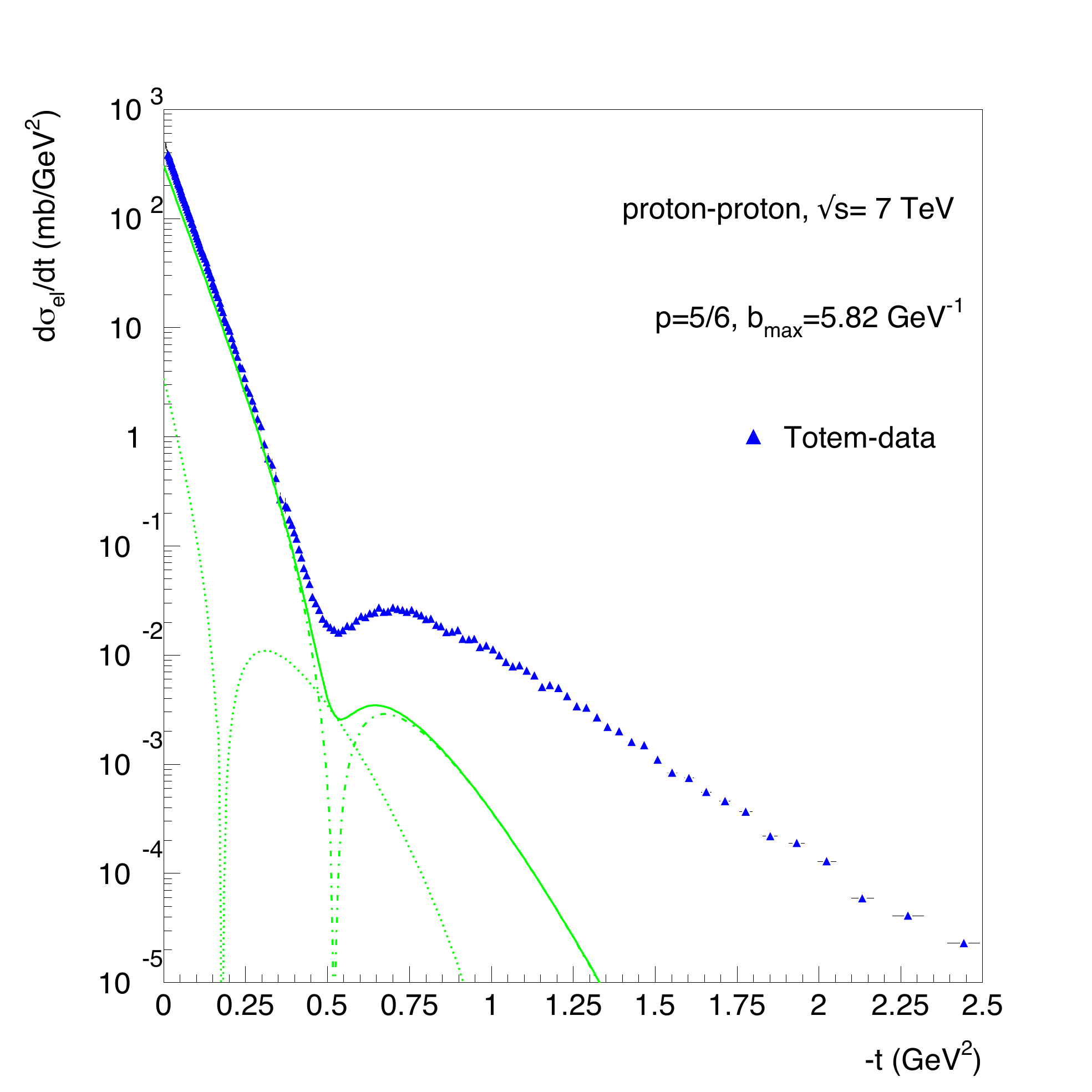}
 \ing[height=8cm,width=8cm]{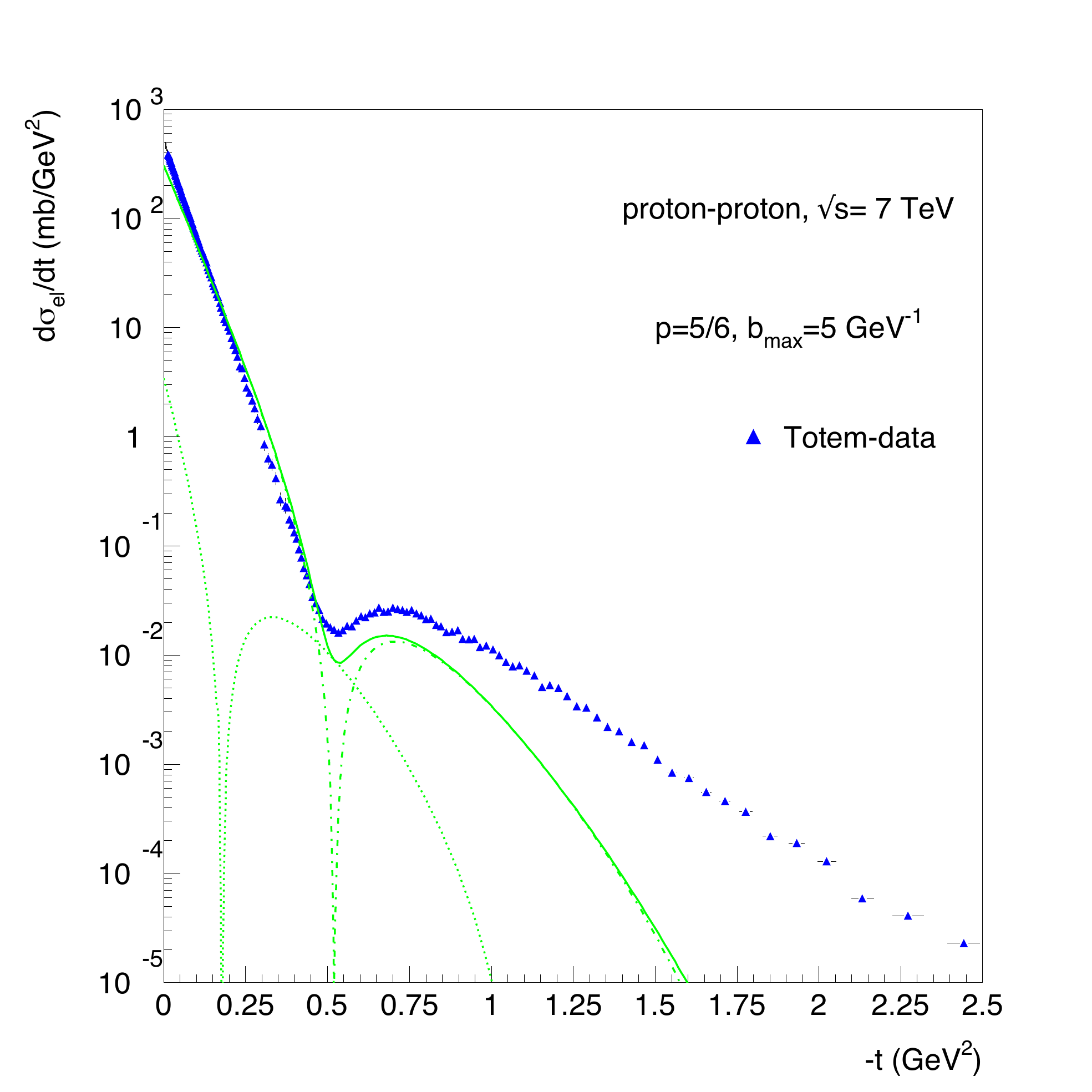}
 \caption{The result of applying the rescattering factor 
  to the one-channel BN model, as described in the text. Bottom figures use GRV densities.}
\label{figs:bn-bmax}
 \eef

In particular:
\begin{itemize}
\item
The two top  figures are   obtained with  MRST72 densities, the top left corresponding to   
the parameter set which reproduces the total cross-section { at LHC7}, MRST72 PDF, $p=0.66,\  p_{tmin}=1.25\ GeV$ \cite{our2011}. 
For the top left plot, we use the $p$ value which reproduces the optical point and  the  slope is obtained from  
the $b_{max}(p=0.66)$ value   consistent with the behavior of $\barn$ for the corresponding densities. However, 
  the position of the dip is still too much to the left. 
To obtain a curve with the dip in the right position one needs an amplitude with a { larger} $p$ value, 
as in the right top plot,
 where  the value for $b_{max}$ is chosen  so as to reproduce the dip. 
 The description of   $\sigma_{elastic}$ with these densities is obtained with  
 $p=0.8$, but then the optical point is lost. 
\item  the bottom  figures are an exercise in showing  
how  the slope $b_{max}$ influences the  depth of the dip: once a $p$-value is chosen  so as obtain the correct dip position, by varying 
$b_{max}$ as a  free parameter  the curve can be lowered and the normalization at the dip 
can be correctly reproduced.
\item The  two bottom figures are obtained with a different PDF set, GRV, and 
parameter $p=5/6$, the value which would  reproduce $\sigma_{elastic}$ with these densities: 
this allows to keep the position of the dip in the proper place. { Lowering $b_{max}$ to a value $b_{max}= 4.75\ GeV^{-1}$ will fully reproduce the dip with these densities.}
\end{itemize}
In all these figures, except for the top left one, the optical point is {\it not} reproduced. The above confirms the well known fact that there is no geometrical scaling at these energies \cite{brogueira}, namely the differential elastic cross-section does not scale with the variable $\tau= -t\ \sigma_{total}$. In particular the position of the dip is {\it not} given by $\tau_{dip}=-t_{dip} \ \sigma_{total}$. Until the BD limit of ${\cal R}_{el}=1/2$, the problem will have  two scales, the total {\it and} the elastic cross-sections. In  the empirical model of \cite{our2013} it appears that this limit, if reached at all, may still be very far and in \cite{Pancheri:2014} we have proposed a different scaling behaviour for the position of the dip \cite{newgs}, namely 
\be
\tau_{dip}=-t\ \sqrt{\sigma_{tot} \sigma_{el}}=constant.
\ee
How to include this in a one channel eikonal model such as the above BN model, is under study.
{Notice that only  recently we became aware of two previous works by Kawasaki, Maehara and Yonezawa \cite{newgs} on the generalized form of geometrical scaling (GS)   in elastic scattering at high energies.} 
 
{Here we mention that a possible way to reconcile with the original GS is to account for the energy evolution of  $\mathcal{R}_{el}(s)$, introducing a saturating function at $s\rightarrow \infty$ \cite{fm2012}, as done in \cite{jdd2013}, for instance.}

\section{Conclusions}
All of the above shows that in a one channel eikonal the position of the dip is inconsistent with the correct 
reproduction of the optical point. This is a well known fact, but we have  discussed  it in detail through 
our mini-jet with soft gluon resummation model, the BN model. 
It also corresponds to the fact that  
there is {\it no} geometrical scaling (at least for the energies probed as yet): $-t_{dip}\ \sigma_{total}$ is not a constant. 
We have discussed this point in a recent note \cite{Pancheri:2014} and refer the reader to the 
considerations exposed in that paper.

In addition to 
the above problems alluded to with this one-channel eikonal, the tail after the dip-bump region is 
also not reproduced at all. This clearly implies that a second term needs to be added, as shown  
through the results of our empirical model in \cite{our2013}, where  we have described data for the elastic 
differential cross-section through two terms, one corresponding {\bf only} to $C=+1$ exchange and 
{ a second one which includes  } $C=-1$ exchange, which is not present here. According to Donnachie and Landshoff, the behavior after  
the dip  arises because of a perturbative contribution from three gluon exchanges, a $C=-1$ term \cite{DL2013}. 
Work is in progress to  examine such a possibility in the context of this BN model.

\end{document}